\documentclass{kluwer}
\usepackage{graphicx}
\begin{document}
\begin{article}
\begin{opening}
\title{$\alpha$-process elements in the Galaxy: a possible GAIA contribution}

\author{Gra\v{z}ina \surname{Tautvai\v{s}ien\.{e}}}
\institute{Institute of Theoretical Physics and Astronomy, Go\v stauto 12,
Vilnius 2600, Lithuania}
\author{Bengt \surname{Edvardsson}}
\institute{Department of Astronomy and Space Physics, Uppsala Astronomical
Observatory, Box 515, SE-751\thinspace 20, Uppsala, Sweden}

\runningtitle{$\alpha$-process elements in the Galaxy: a possible GAIA contribution}
\runningauthor{G. Tautvai\v{s}ien\.{e} and B. Edvardsson}

\begin{abstract}
The sensitivity of stellar spectra to $\alpha$/Fe abundance changes is investigated 
with the aim to be detected photometricly and employed for
scientific goals of the GAIA mission. A grid of plane parallel, line blanketed,
flux constant, LTE model atmospheres with different [$\alpha$/Fe] ratios was
calculated. As a first step, the modelled stellar energy fluxes for solar-type stars 
and giants were computed and intercompared.
The spectral sensitivity to $\alpha$/Fe abundance changes
is noticeable and has to be taken into account when selecting photometric
filters for GAIA.
The Ca\,{\sc ii} H and K lines and Mg\,{\sc i}\,b triplet are
the most sensitive direct indicators  of $\alpha$/Fe abundance changes. 
\end{abstract}

\keywords{Galaxy evolution, photometric techniques, orbiting observatory GAIA}



\end{opening}
\section{Introduction}

The evolution of a galaxy is closely related with a gradual chemical
enrichment. The enrichment and spatial distribution of the chemical elements
depend on various galactic and stellar processes. In particular, the star
formation (SF) history, the time delay between SF and the enrichment of the
interstellar medium (ISM), the metal dependency of the nucleosynthesis, the
galactic gas flows, and the mixing processes in the ISM are important. Since
the individual elements are produced at various sites and on different time
scales, the observed abundances are very useful in describing galactic
evolution. The main archeological tracers of the chemical evolution are the
elements produced
on a short time scale ($10^7$ years) by massive stars ending as core-collapse
supernovae (here called SN~II) and on a longer time scale ($10^9$ years)
by Type~Ia (SN~Ia) supernova events.
SNe~II contribute to the enrichment of the
interstellar medium mainly with elements produced by the capture of
$\alpha$-particles ($\alpha$-elements) and from the $r$-process, and SNe~Ia
predominantly
produce elements belonging to the Fe peak. Consequently, one of basic tools
to constrain the evolution of a galaxy is the analysis of relations
between ratios of [$\alpha$-element/Fe]
and Fe abundances [Fe/H]\footnote
{In this paper we use the customary spectroscopic notation
[X/Y]$\equiv \log_{10}(N_{\rm X}/N_{\rm Y})_{\rm star} -
\log_{10}(N_{\rm X}/N_{\rm Y})_\odot$}
for stars born at different times and
in different parts of a galaxy. E.g., the theoretical evolutionary model of the 
Milky Way 
galaxy recently proposed by Chiappini et al.\ (2001) predicts
a slight decrease with distance in the average [$\alpha$/Fe] ratios in
stars born in the Galactocentric distance range 4--10 kpc and an increase
with distance of this ratio in the range 10--18 kpc.

A first glance at the temporal behavior of $\alpha$-elements shows that
most of the metal-poor stars in the Galaxy appear to have been formed with enhanced
abundances of oxygen and other $\alpha$-elements (i.e., Ne, Mg, Si, S, Ar,
Ca and Ti). For stars with [Fe/H]$\le-1$, the mean value of [$\alpha$/Fe] lies
between $+0.3$ and $+0.4$, with no discernible dependence on metallicity
(cf. Pagel and Tautvaisiene 1995; Samland 1998).
A more precise analysis shows that there is a significant population
of field stars with [$\alpha$/Fe]$\approx 0.0$ (see Fig.4 by Nissen and
Shuster, 1997).
Surveys by Shuster et al.\ (1993) and Carney et al.\ (1996) report evidence for
about $0.1-0.2$\,dex variations in the [$\alpha$/Fe] value at a fixed [Fe/H].
Carney et al.\ (1997) have found that the high-velocity subgiant with
apogalacticon distance over 20 kpc BD\,$+80^{\circ} 245$ has
$<[\alpha/{\rm Fe}]>=-0.29\pm 0.02$ despite to its low metallicity, [Fe/H]=$-1.86$.
At the same time there are metal deficient giants with [$\alpha$/Fe] of +0.7\,dex
as reported by Giridhar et al.\ (2001).
It is important to map the abundance pattern of $\alpha$-elements in the Galaxy
and understand the origin of variations.

A central element of the GAIA mission is the determination of the star formation
histories, as described by the temporal evolution of the star formation rate, and
the cumulative numbers of stars formed in the bulge, inner disk, Solar
neighbourhood, outer disk and halo of our Galaxy (Perryman et al.,\ 2001).
Such information, together with the kinematic information from GAIA, and
complementary chemical abundance information, again primarily from GAIA,  
may give us the full evolutionary history of the Galaxy.
Knowledge of information on $\alpha$/Fe abundance ratious in stars is very important
for their age determination (cf. VandenBerg et al., 2000; Salasnich et al.,
2000). In our study, a first
attempt is made to investigate the sensitivity of stellar spectra to $\alpha$/Fe
abundance variations and their detection by GAIA photometry.

\section{Model atmospheres with enhanced $\alpha$/Fe ratios }

A grid of plane parallel, line blanketed, flux constant, LTE model
atmospheres with enhanced  $\alpha$/Fe ratios was calculated
with the updated version of the MARCS code (Gustafsson et al., 1975)
using continuous opacities from Asplund et al.\ (1997) and including UV
line blanketing as described by Edvardsson et al.\ (1993).
The grid contains model atmospheres with effective temperatures
$4500\,{\rm K}\le T_{\rm eff}\le 6500{\rm K}$,
$1.5\le \log g\le 4.5$, $-2\le {\rm [Fe/H]}\le 0$ and
$0.0\le$\,[$\alpha$/Fe]\,$\le 0.4$.
By $\alpha$ elements we here mean O, Ne, Mg, Si, S, Ar, Ca and Ti.

\section{Sensitivity of stellar spectra to  $\alpha$/Fe changes}

\begin{figure}
\centerline{\includegraphics[width=34pc]{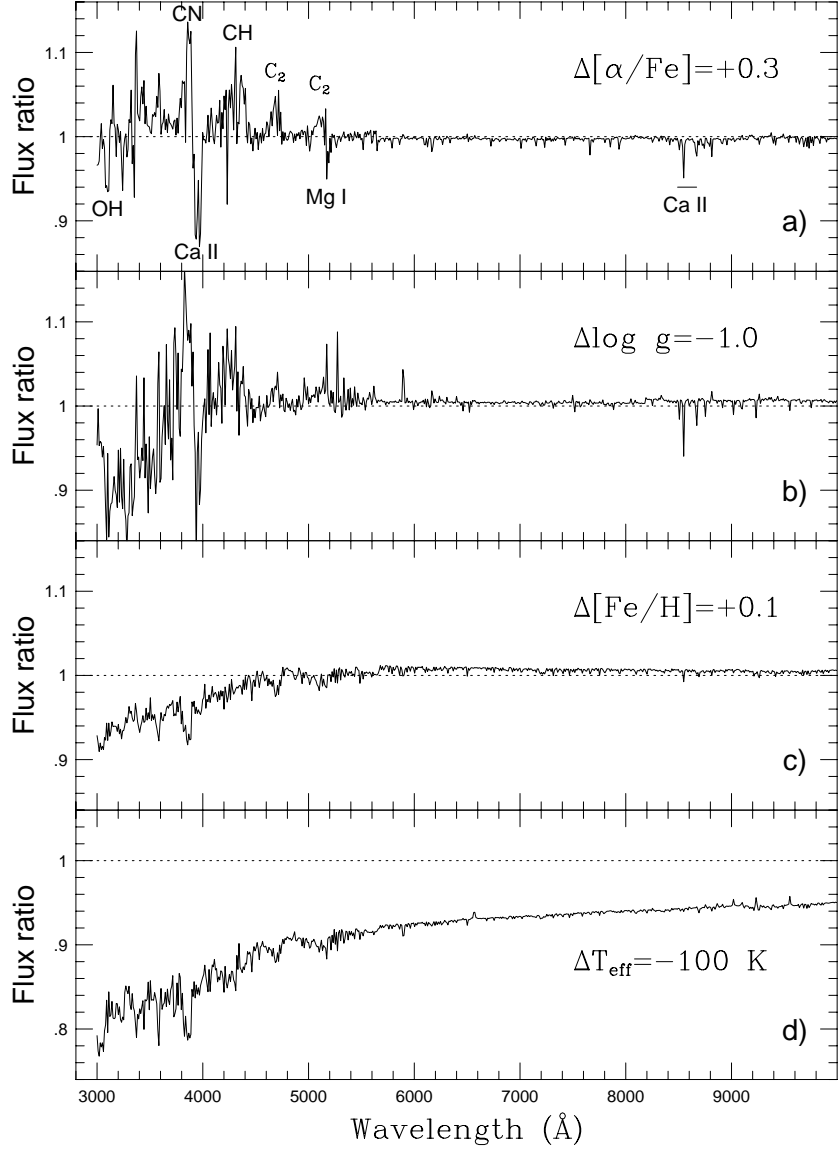}}
\caption{
Sensitivities of the modelled surface energy flux ratios to variations in 
fundamental parameters and chemical abundances.
The standard model has $T_{\rm eff}=5500$\,K, $\log g=4.0$, [Fe/H]\,$=-0.4$
and [$\alpha$/Fe]\,=0.0. See Sect.\,3 for more explanations.
}
\end{figure}

\begin{figure}
\centerline{\includegraphics[width=34pc]{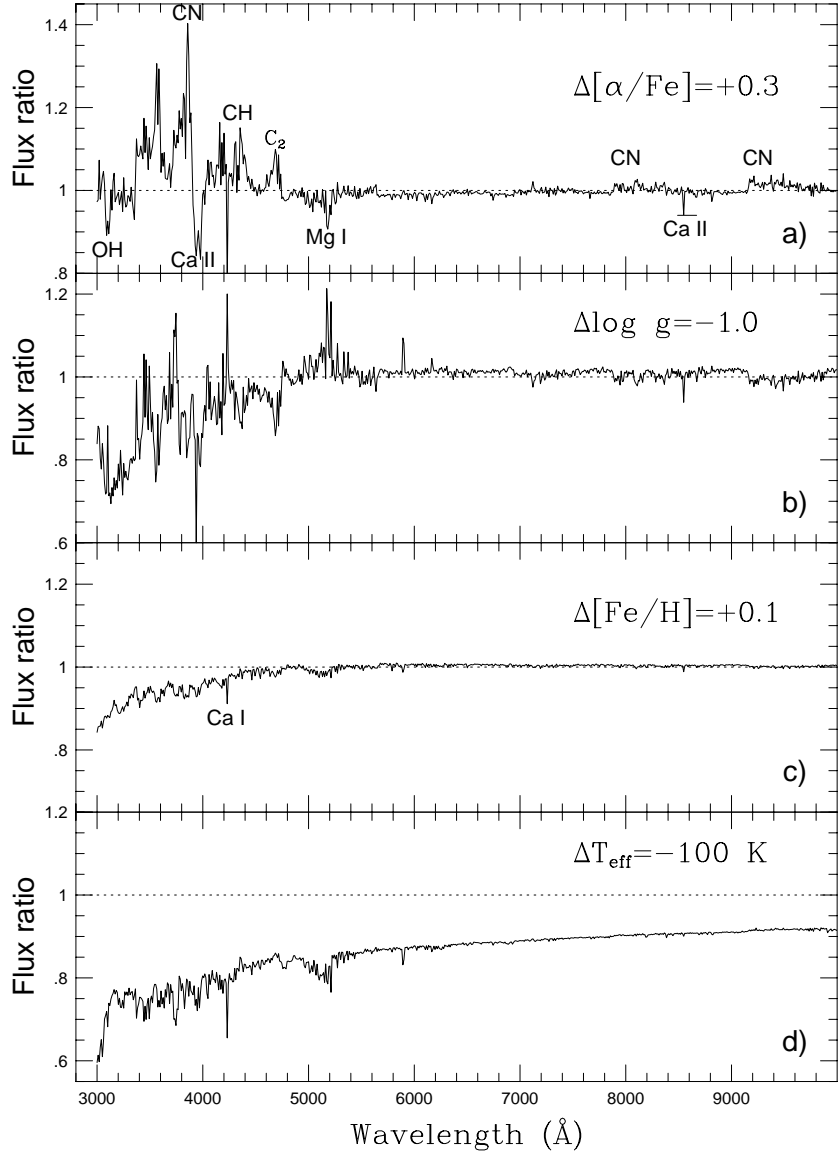}}
\caption{
Sensitivities of the modelled surface energy flux ratios to variations in 
fundamental parameters and chemical abundances.
The standard model has $T_{\rm eff}=4500$\,K, $\log g=3.0$, [Fe/H]\,$=-0.4$
and [$\alpha$/Fe]\,=0.0. Notice that the vertical scaling differs from that 
in Fig.~1.
}
\end{figure}

\begin{figure}
\centerline{\includegraphics[width=27pc]{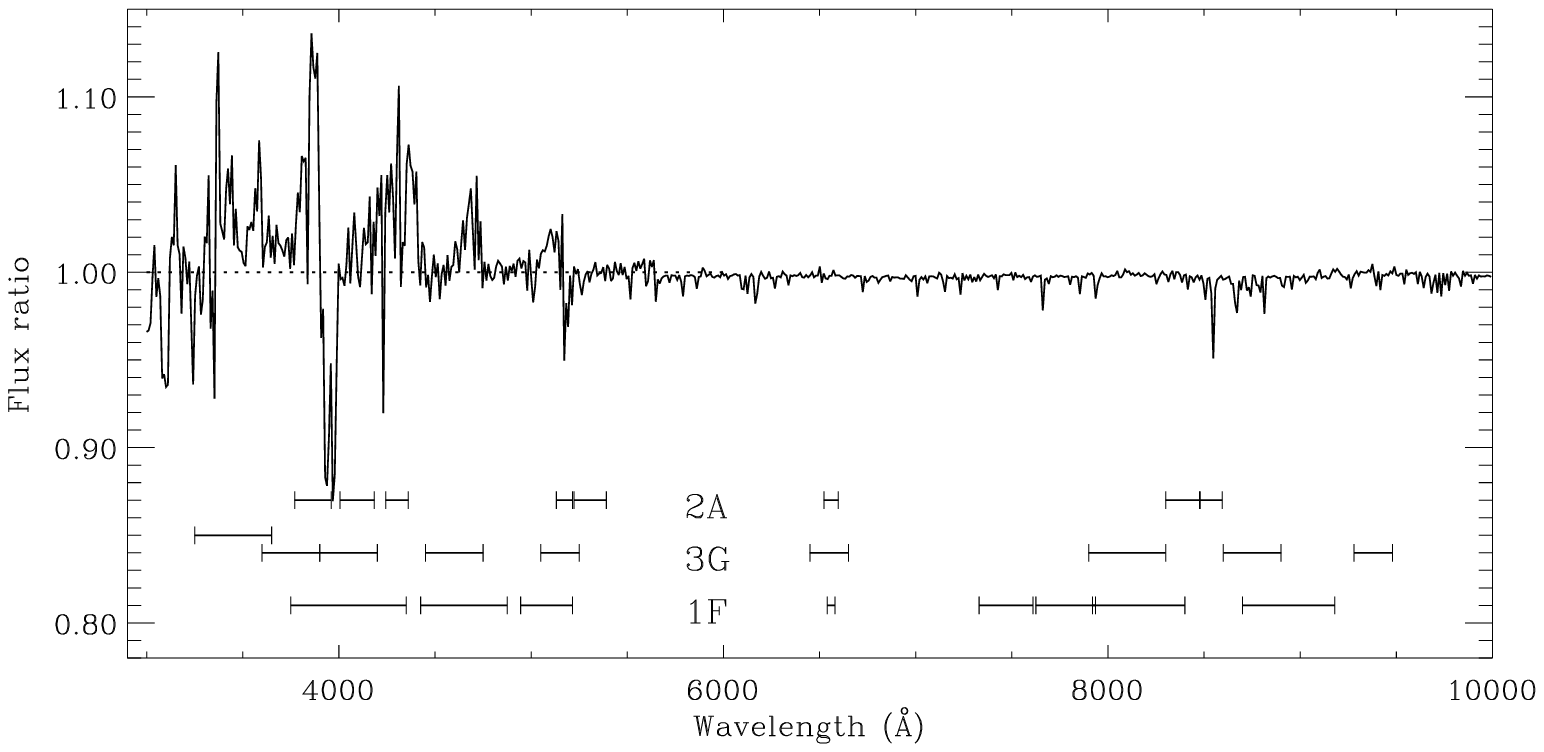}}
\caption{
The ratio of modelled surface enery fluxes with $\Delta [\alpha /{\rm Fe}]=+0.3$ for 
$T_{\rm eff}=5500$\,K, $\log g=4.0$ and [Fe/H]\,$=-0.4$, plotted 
together with the three photometric systems proposed
for GAIA. See Sect.\,3 for more explanations.
}
\end{figure}

In order to see what spectral regions are most sensitive to $\alpha$/Fe
changes, as a first step, we computed stellar surface fluxes with a wavelength
sampling of $R\approx 20\thinspace 000$ for solar type stars
and giants and investigated their ratios. E.g. Fig.1$a$ shows the smoothed
ratio of the flux distributions with $T_{\rm eff}=5500$\,K, $\log g=4.0$,
[Fe/H]$=-0.4$ and two different [$\alpha$/Fe] ratios: 0.3 and 0.0\,dex. 
The approximate difference of 0.3\,dex was found between
thin- and thick-disk stars in the solar vicinity (cf. Fuhrmann, 1998; 
Tautvai\v{s}ien\.{e} et al., 2001).
The major wavelength features directly sensitive to $\alpha$/Fe
in dwarf stars are the Ca\,{\sc ii} H and K lines and the IR triplet,
the OH bands around 3100\,\AA\
and also the Mg\,{\sc i}\,b triplet region.
The latter region is also affected by MgH molecular bands in cool stars.
These features are indicated below the curve in panel $a$.
There is also an indirect effect of the $\alpha$-element abundances which
is seen on the carbon molecular bands: the higher oxygen abundance binds more
free carbon into CO, which weakens other carbon molecular bands.
The strongest examples are indicated above the curve in panel $a$.
It is obvious that these secondary effects are quite dramatic and
should be taken into account while using carbon features for photometry.

In Fig.1$b$, the effect of decreasing the surface gravity by 1.0\,dex is shown.
The effects are qualitatively similar to those of the $\alpha$-element abundance 
increase
in several wavelength regions. A clear difference starts only bluewards from 
3800\,\AA. The strong gravity effect on the Ca\,{\sc ii} lines make their use 
for $\alpha$-element determinations quite dependent on a precise $\log g$
determination. It is interesting to notice the opposite gravity effects on the
Ca\,{\sc ii} and the Mg\,{\sc i} features.
The former are radiation damped (thus not pressure-sensitive) and strengthened
by the decreasing H$^-$ continuous opacity and increasing degree of ionization,
while the latter are pressure-broadened and thus weakened both by the weaker
gas pressure and by the higher degree of ionization.
The sensitivity of the spectrum to changes in overall metallicity and effective
temperature are shown in Figs.\,1$c$ and 1$d$, respectively. 
The prominent feature of CN near 3850\,\AA\ is dependent on very many parameters,
including nitrogen abundances, which are known to vary during stellar evolution.

Fig.\,2 shows sensitivities of the modelled surface flux ratios to variations in 
fundamental parameters and chemical abundances for a giant star with 
the standard model of $T_{\rm eff}=4500$\,K, $\log g=3.0$,
[Fe/H]$=-0.4$ and [$\alpha$/Fe]=0.0. As it is seen from the vertical scales, 
in giants the effects of [$\alpha$/Fe] abundance changes and other parameter 
variations are considerably larger than for the dwarfs.  
 
In Fig.\,3 we display the three medium-band photometric systems proposed for GAIA:
2A by Munari (1998), 3G by H{\o}g et al.\ (2000), and 1F by Grenon et
al.\ (1999, the wide filters F33, F57 and F67 are not displayed) 
along with the modelled ratio of surface energy fluxes with
$\Delta [\alpha /{\rm Fe}]=+0.3$ for  $T_{\rm eff}=5500$\,K, $\log g=4.0$ and
[Fe/H]\,$=-0.4$. 
Figs.\,1 -- 3 show that the filter centered at 3450\,\AA\ in the 3G system
is very useful in determining surface gravities, the filter centered at 3750\,\AA\
may evaluate the CN feature, while the filter centered at 4050\,\AA\ --
the H and K lines. However, the filter at 5150\,\AA\
includes both C$_2$, MgH and Mg\,{\sc i}\,b lines which may cause
a confusion. Here the filter of the 2A system  centered exactly on the
Mg\,{\sc i}\,b triplet seems to be better. 
Filters of the 1F system are quite broad and may cause difficulties in accounting 
for $\alpha$-element variations. 
One filter in the GAIA system could be set on the  Ca\,{\sc ii} IR triplet as well.

The first qualitative investigation of sensitivity of stellar spectra to 
$\alpha$/Fe abundance variations, presented in this study, indicates that 
the spectral sensitivity to $\alpha$/Fe abundance changes is noticeable. 
It has both direct and indirect influence to stellar spectra.
A possibility to employ the Ca\,{\sc ii} H and K lines and Mg\,{\sc i}\,b triplet 
might be considered
for the photometric determination of $\alpha$/Fe abundance ratios with GAIA photometry. 
The examples for the dwarf and giant stars with $\Delta [\alpha$/Fe]=0.3, 
displayed in Figs.~1 and 2, show that 
filters of about 80--100~\AA\ width centered on these features could be used. 
In the interval of the spectrum 3905--4005~\AA\ with the Ca\,{\sc ii} lines, the 
intensity of the spectrum drops down by 0.08~mag in the dwarf and by 0.11~mag in the 
giant. In the interval 5160--5240~\AA\ with the Mg\,{\sc i}\,b triplet, the intensity 
of the spectrum drops by 0.02~mag and 0.05~mag, respectively. 
Assuming the end-mission photometric GAIA accuracy and three slots of filters 
(ESA, 2000), 
from the  Ca\,{\sc ii} lines, the accuracy of [$\alpha$/Fe]$=\pm 0.1$ 
might be preserved for the giant (with parameters under consideration) down to about 
17.0~mag and for the dwarf down to  
about 16.3~mag. From the Mg\,{\sc i}\,b triplet, the same accuracy of $\pm 0.1$~dex 
might be preserved for the giant down to about 17.3~mag and for the dwarf down to 
about 15.6~mag. In case the accuracy of $\pm 0.2$~dex is also acceptable, the 
stars of 17.9 and 16.7~mag could be investigated using the Ca\,{\sc ii} lines and 
stars of 18.2 and 16.5~mag using the Mg\,{\sc i}\,b triplet, correspondingly.          
 
The work presented in this paper marks the beginning of a large work to be done 
in preparations for the photometric investigation of $\alpha$-elements in the 
Galaxy. 
Under the assumption of known effective temperature, metallicity and
surface gravity, the carbon, nitrogen  and $\alpha$-element abundances might be
determined by means of photometric indices as well.  

\section{Conclusions}

The spectral flux sensitivity to $\alpha$/Fe abundance changes is noticeable 
and has to be taken into account in GAIA photometry. 
The Ca\,{\sc ii} H and K lines and Mg\,{\sc i}\,b triplet are
most sensitive direct indicators of $\alpha$/Fe abundance changes and 
might be used for the photometric determination of $\alpha$-element abundances.
The photometric systems proposed for GAIA have to be carefully tested for 
accounting of the effects of $\alpha$-element abundance variations and their 
determination.

Photometric classification of stars should provide as many physical parameters
as possible. Depending on the accuracy with which the fundamental parameters
are known, we should seek to determine abundances
not only of $\alpha$-elements but of carbon and nitrogen as well.

\acknowledgements
We wish to thank Vytautas Strai\v{z}ys, Michel Grenon and Vladas Vansevi\v{c}ius
for helpful discussion. G.T. acknowledges support from the Nordic Research
Academy (REF NB00-N030) and NATO Linkage grant CRG.LG 972172.
B.E. acknowledges support by the Swedish Natural Sciences Research Council (NFR).

{}
\end{article}
\end{document}